\documentstyle[epsf,epsfig,graphicx]{elsart}
\begin{document}
\begin{frontmatter}
\newcommand{\postscript}[2]{\setlength{\epsfxsize}{#2\hsize}
\centerline{\epsfbox{#1}}}
\title{Neutron$-^{19}$C scattering near an Efimov state}
\author[itapeva]{M.T. Yamashita}\footnote{Email: yamashita@itapeva.unesp.br},
\author[ITA]{T. Frederico}\footnote{Email: tobias@ita.br} and
\author[IFT]{Lauro Tomio}\footnote{Email: tomio@ift.unesp.br}
\address[itapeva]{Campus Experimental de Itapeva, Universidade Estadual 
Paulista,\\ 18409-010, Itapeva, SP, Brazil.}
\address[ITA]{Instituto Tecnol\'ogico de Aeron\'autica,
CTA,\\ 12228-900, S\~ao Jos\'e dos Campos, Brazil.}
\address[IFT]{Instituto de F\'\i sica Te\'orica, 
Universidade Estadual Paulista,\\ Rua Pamplona, 145, 01405-900, S\~{a}o Paulo, Brazil.}
\date{\today}
\maketitle
\begin{abstract}
The low-energy neutron$-^{19}$C scattering in a neutron-neutron-core
model is studied with large scattering lengths near the conditions
for the appearance of an Efimov state. We show that the real part of
the elastic $s-$wave phase-shift ($\delta_0^R$) presents a zero, or a
pole in $ k\cot\delta_0^{R}$, when the system has an Efimov
excited or virtual state. More precisely the pole scales with the
energy of the Efimov state (bound or virtual).  We perform
calculations in the limit of large scattering lengths, disregarding
the interaction range, within a renormalized zero-range approach
using subtracted equations. It is also presented a brief discussion
of these findings in the context of ultracold atom physics with
tunable scattering lengths.
\newline\newline
PACS {03.65.Nk, 11.80.Jy, 03.65.Ge, 21.45.-v, 21.10.Dr}
\end{abstract}
\end{frontmatter}
The experimental observation of an Efimov resonant state~\cite{Efimov} in an 
ultracold gas of Cesium with tunable interactions, performed by the Innsbruck 
group~\cite{kraemer}, evidenced the universal properties of large three-body 
quantum states~\cite{amorim,bedaque-bira,rmpjensen,braaten}.
The observations of \cite{kraemer} gave support to several theoretical analysis 
on the possibility to identify Efimov states considering precise determinations 
of three-body recombination rates~\cite{3brecomb}. They were also consistent with 
calculations of resonances for the atom-dimer channel~\cite{ADreson} and for 
continuum triatomic Borromean configurations~\cite{resonance}.
In this situation, the properties of these large systems are critically dependent 
on the physical scales of the two-body subsystems, fixed by the two-body 
scattering lengths and one three-body scale~\cite{danilov}. 
See also the discussions presented in Refs.~\cite{adh88,adh95} on the scales 
and the relation between Efimov effect and the Thomas collapse of the three-body 
ground-state energy~\cite{thomas}.
 
The appearance of resonant three-body states in the maximally symmetric 
state is controlled by ratios of the two- and three-body scales~\cite{resonance}. 
For example, the freedom to change the two-body energy ($E_2$), as in
ultracold trapped gases with tunable interactions, allowed the study
of the behavior of an Efimov state energy ($E_3$) when  $E_2$ is
changed. For three identical bosons, $E_3$ follows the route
virtual-bound-resonance for a large two-body scattering length
varying from positive to negative values passing through the
infinite (this corresponds to a change from a bound to a virtual
state in the two-boson system)~\cite{fb18}.

A bi-dimensional map in the parametric space can be defined in the
Efimov limit by the critical conditions for an excited state in
three-body systems  with two-identical particles. The scattering
lengths of the two-body subsystems and one three-body scale define
the parametric space. The border of the map encloses a region where
excited states do exist (see Refs. \cite{amorim} and \cite{hammer08}). 
Crossing the critical boundary implies that an excited state becomes 
a virtual one, when at least one subsystem is bound~\cite{c20}, or a 
continuum resonance in the Borromean case. These qualitative features 
are found to be independent on the mass ratios~\cite{c20}.

The Borromean case of a three-boson continuum resonance was recently
evidenced \cite{kraemer} in an experiment with trapped ultracold
cesium gas near a Feshbach resonance. It is hoped that mixtures of
different-mass atoms in traps with tunable interactions allow to
check the transition from Borromean to non-Borromean situations
where a continuum resonance becomes bound and turns to a virtual
state by changing the sign of the scattering length.

In the nuclear context the universal physics associated with the
Efimov phenomenon is hoped to be evident in light halo nuclei.
Recently it was suggested that the  $n-n-^{18}$C ($n$ represents a
neutron) system would exhibit low-energy resonant properties
reminiscent of an Efimov state \cite{ArPRC04}. The study of the halo
nucleus $^{20}$C  modeled as $n-n-^{18}$C system with an $s-$wave
short-range interaction between the pairs can be a playground to
search for evidences of universal three-body physics. Using a
zero-range interaction we have shown that this three-body halo
system presents a virtual state that turns into an excited state
when the $^{19}C$ binding is decreased \cite{c20}. Close to this
condition, it is suggestive to study the low-energy $n-^{19}$C
elastic scattering in $s-$wave as the Efimov poles of the scattering
amplitude are near the elastic scattering threshold.

Indeed, in the neutron-deuteron ($n-d$) doublet $s-$wave elastic
scattering, the energy dependence of the phase-shift has to be
changed from a simple effective range formula to a more complex one
\cite{delves,vanoers}. The change in the off-shell behavior of the 
two-body potential, or three-body forces, modifies the corresponding
phase-shift correlated to the triton binding in a way that the
scattering length can vanish. This was already seen in the
well-known Phillips plot where the doublet scattering length is
presented as a function of the triton binding energy
\cite{phillips}. In the case of zero scattering length, the
function $k\cot\delta_0$ has a pole at zero relative $n-d$ kinetic energy,
where $k$ is the on-energy-shell incoming and final relative momentum
related to the energy of the three-body system. 
The necessity of a pole in $k\cot\delta_0$  has been pointed out   in
the analysis  of the experimental data for the $n-d$
doublet $s-$wave phase-shift \cite{vanoers}. The analytical structure
of $k\cot\delta_0$ in the $n-d$ elastic doublet state scattering has
been also studied in several
works~\cite{reiner,wfuda,gfuda,adhikaritorr,simenog87}. Van Oers and
Seagrave proposed to incorporate a pole in the phenomenological
effective range formula used to fit the $k\cot\delta_0$ low energy
data for the $n-d$ $s-$wave doublet state just below the elastic
threshold. The effective range expansion has a form given by
\cite{vanoers}
\begin{equation}
 k\cot\delta_0 = -A + B k^2 - \frac{C}{1+D k^2}, \label{kcotnd}
\end{equation}
where $A$, $B$, $C$, and $D$ are fitted constants. Its position and
residue have been calculated from dispersion relations as well as
exact solutions of three particle equations with separable
interactions by Whiting and Fuda \cite{wfuda}. Later on the
existence of the triton virtual state was found on the basis of the
effective range expansion \cite{gfuda}. After that, it was suggested
from the solution of a three-body model with separable two-body
interactions that the triton virtual state appears from an excited
Efimov state moving to the non-physical energy sheet through the
elastic cut \cite{adhprc821}. The same phenomenon is found in the
case of the excited Efimov state of the two-neutron halo of
$^{20}$C, where the pole of the S-matrix migrates to the second
energy sheet through the elastic $n-^{19}$C cut when the binding
energy of the neutron in $^{19}$C is increased.

The understanding of the low-energy properties of $n-n-~^{18}$C
system also requires the study of the behavior of the scattering
amplitude when an Efimov state is near the physical region. That
state appears either  as a bound excited state or as a virtual state
depending on the low-energy parameters of the two-body subsystems as
well the three-body scale (e.g., the ground state energy of
$^{20}$C). As we have shown \cite{c20}, the mass ratios are not
relevant for the trajectory of an Efimov excited state turning into
a virtual state with the increase of the $^{19}$C binding energy, in
correspondence to what has been found in the case of the trinucleon
system \cite{adhprc821}. Our next task is to find if the $n-^{19}$C
$s-$wave low-energy phase-shift exhibits the same analytical
properties as found in the case of the $n-d$ doublet state.  It is
interesting to find if exists a low-energy pole in
$ k\cot\delta_0^{R}$ and to check if that property results to be
independent on the mass ratios. This is suggested by the $n-^{19}$C
and $n-d$  similar  cut structure of the scattering amplitudes, which
is already implied by the very same behavior of the excited state
turning a virtual Efimov state. We also note that the expression
(\ref{kcotnd}) was used to fit the real part of the phase-shift.

To support the possible qualitative and universal common behavior of
the elastic $s-$wave scattering amplitudes of $n-^{19}$C and the
$n-d$ doublet state, let us discuss the example of the atom-dimer
scattering length as a function of the atom-atom scattering length
$a$ and a three-body scale, namely $\Lambda_*$, as considered in
Ref.~\cite{braaten}. In the case of three bosons the atom-dimer 
scattering length is approximately given by
\begin{eqnarray}
a_{AD} = ( 1.46 - 2.15 \tan[s_0 \ln(a\Lambda_*) + 0.09] ) a \ ,
\label{bra1}
\end{eqnarray}
where $s_0 = 1.00624$. The atom-dimer scattering   length is a
solution of a one-body problem in the attractive long-range
$\rho^{-2}$ potential responsible for the Efimov states. 
The dimensionless product $a\Lambda^*$ controls the number of
three-body bound states, with $a\Lambda^*\rightarrow \infty$ implying in
an infinite number of Efimov states. The above limit can be
realized either by $a\rightarrow \infty$ for a fixed $\Lambda^*$ or
$\Lambda^*\rightarrow \infty$ for a fixed $a$. 
Efimov states correspond to the first case, whereas the 
Thomas collapsed states~\cite{thomas} appear in the second case. 
So, the formation of bound states depending on the
dimensionless product $a\Lambda^*$ relates the Thomas collapse and
Efimov effect \cite{adh88}.
The expression (\ref{bra1}) allows a zero scattering length or
a pole of the effective range expansion at zero energy at a value
$a=a_0$ given by 
$a_0= 1.6543/\Lambda_*$.
It is worthwhile to stress that (\ref{bra1}) also allows
$a_{AD}\rightarrow \pm\infty$, or an  Efimov state at the threshold,
for a value of the atom-atom scattering length $a = a_B  = 4.3563/\Lambda_*$,
obtained from $
s_0 ln(a_B\Lambda_*) + 0.09={\pi}/{2} .$
So, by changing the two-body scattering length
from $a=a_0$ to $a=a_B$,  a bound Efimov state is produced at the
threshold. In order to change $a_{AD}$ from zero to an infinite
value the ratio of atom-atom scattering lengths can be estimated
from the above as given by
\begin{eqnarray}
\frac{a_B}{a_0}={\rm exp}\left({\frac{{\pi}/{2}-0.59654}{s_0}}\right)
\simeq 2.633.
\label{bra3}
\end{eqnarray}
Therefore, from our qualitative discussion we observe that the
physics related to the Efimov effect is also implying a zero in the
atom-dimer scattering length, and consequently a pole in the
effective range expansion at zero kinetic energy. Our next
calculations will substantiate that a pole in the effective range
expansion of the $n-^{19}$C elastic phase-shift appears in a quite
good qualitative agreement with the above. In such a case, the
three-body system has unequal mass particles, with $m_1=m_2$
(neutron mass) and $m_3/m_1=A=18$, implying in a corresponding
variation of $s_0$. Extracted from Fig.~52 of \cite{braaten}, one
obtains $s_0\sim 1.12$ for $A=18$, such that ${a_0}/{a_B}\sim
0.419$. This result will be shown to be quite close  to our
numerical results.

The aim of this work is to study the low-energy $s-$wave
$n-^{19}$C scattering in a neutron-neutron-core model near the
critical condition for the appearance of an excited Efimov state. In
this situation it is reasonable to use an $s-$wave zero-range
interaction between the pairs, with a three-body scale tuned to the
separation energy of the two halo neutrons of $^{20}$C. The
three-body scale is introduced through a subtraction in the kernel
of the integral equations ~\cite{adh95}. Alternatively, another
subtraction scheme also allows the introduction of the physical
information of the three-body system at the elastic scattering
threshold, see e.g. \cite{afnan}.
We will present results for the $s-$wave phase-shift ($\delta_0$) of
the $n-~^{19}C$ elastic scattering for a fixed $^{20}$C energy and
singlet scattering length, for different values of the neutron
separation energy in $^{19}$C, from 200 up to 850 keV. That range is
chosen to study how $\delta_0$ behaves as the excited Efimov state
moves into the unphysical energy sheet turning to a virtual state.
We will show that the real part of $\delta_0$ presents a zero, or a
pole in $ k\cot\delta_0^{R}$. That pole scales with the energy of the
Efimov state (bound or virtual). Our goal is to study the low-energy
phase-shift in the limit of large scattering lengths where
corrections due to the interaction range are not relevant for the
qualitative properties, dominated by a universal scaling behavior
(see e.g. \cite{rmpjensen} and \cite{braaten}). We will also present
a brief discussion on how to apply our findings, through
universality,  to the context of trapped ultracold atomic systems with
different species near a Feshbach resonance.

Next, we present the basic formalism to calculate the
$n-^{19}$C elastic scattering amplitude. The input of our
subtracted scattering equations are the $n-n$ virtual state energy
that will be fixed at $E_{nn}= -$143 keV, the energy of the halo
neutron bound in $^{19}$C, $E_{nc}$, and the $^{20}$C that has a
ground state energy of $-$3.5 MeV. The neutron separation energy in
$^{19}$C has a sizable error with value of $-$160$\pm$110
keV~\cite{audi} and $-$530$\pm$130 keV from Ref. \cite{naka99}. In our
study we will allow wide variations in the $n-^{18}$C energy in a
range between 200 up to 850 keV to include the experimental values.

Our units are such that $\hbar=m_n=1$, where $m_n$ is the mass of
the neutron, $m_c = A m_n$ is the core mass ($A=$ the 
core mass-number) and $\mu_{nc}=Am_n/(A+1)$ is the reduced mass of 
the $n-c$ system. In the present case, the core is identified with
$^{18}$C.
In the following formalism, the energies for the two-body ($E_{nn}$ and 
$E_{nc}$) and three-body ($E_3\equiv E$) systems are, respectively, redefined to 
$\varepsilon_{nn}$, $\varepsilon_{nc}$, and $\cal E$, 
with the conversion unit factor, $\hbar^2/m_n=$41.47 MeV fm$^2$, being 
the same for all the energies:
$E_{nn(nc)}=(\hbar^2/m_n)\varepsilon_{nn(nc)}$, 
and $E_3\equiv E=(\hbar^2/m_n){\cal E}$. 
The partial-wave elastic  $n-(nc)$ scattering equation has been
already discussed in detail in Ref. \cite{c20}. 
To make clear to the reader the main ingredients of the model, we briefly 
review the formalism in the following. The spectator function $\chi_n(\vec q)$,
which represents the relative motion between the neutron and
$^{19}$C target brings the boundary condition of the elastic
scattering as:
\begin{equation}
\chi_n(\vec q)\equiv(2\pi)^3\delta(\vec q-\vec k_i)
+4\pi\frac{h_n(\vec q;{\cal E}(k_i))}{q^2-k_i^2-{\rm i}\epsilon} ,
\end{equation}
where $h_n(\vec q;{\cal E}(k_i))$ is the scattering amplitude, and 
$\vec{q}$ is the momentum of the spectator particle ($n$) with respect 
to the center-of-mass (CM) of the other two particles ($n-c$).
The on-energy-shell incoming and final relative momentum are
related to the three-body energy ${\cal E}\equiv 
{\cal E}_i\equiv{\cal E}(k_i)$ by
$k\equiv k_i\equiv |\vec k_i|=|\vec k_f| =
 \sqrt{[2(A+1)/(A+2)]\left({\cal E}_i-\varepsilon_{nc}\right)}.
$ 
The coupled partial-wave scattering equations can be cast in a
single channel Lippmann-Schwinger-type from the relevant amplitude
$h_n^\ell$:
\begin{eqnarray}\label{fnsca}
&& h_n^\ell(q;{\cal E})={\cal V}^\ell(q,k;{\cal E})
+\frac{2}{\pi} \int_0^\infty \hspace{-0.1cm}dp p^2\;
\frac{{\cal V}^\ell(q,p;{\cal E})\;
h_n^\ell(p;{\cal E})}{p^2-k^2-{\rm i}\epsilon} .
\end{eqnarray}
The kernel of the integral equation for the  $n-(nc)$ channel
amplitude contains the contribution of the coupled $(nn)-c$ channel
as given by the integration seen in the expression below:
\begin{eqnarray}
{\cal V}^\ell(q,p;{\cal E})&\equiv&
\pi\frac{(A+1)}{A+2}{\bar\tau_{nc}}(q;{\cal E})\times \nonumber
\\ \times &&
\left[K_2^\ell(q,p;{\cal E})+ \int_0^\infty dp'p'^2
K_1^\ell(q,p';{\cal E})\tau_{nn}(p';{\cal E}) K_1^\ell(p,p';{\cal
E}) \right] ,  \label{Veq}
\end{eqnarray}
where
\begin{eqnarray} \tau_{nn}(q;{\cal E})
&\equiv& \frac{-2}{\pi}\left[ \sqrt{|\varepsilon_{nn}|}+\sqrt{
\frac{A+2} {4A}q^2-{\cal E}} \right]^{-1}, \label{taunn}
\\
{\bar\tau_{nc}}(q;{\cal E})&\equiv& \frac{-1}{\pi}
\left(\frac{A+1}{2A}\right)^{\frac 32}
\hspace{-0.1cm}\left(\hspace{-0.2cm}\sqrt{|\varepsilon_{nc}}|+\hspace{-0.1cm}\sqrt{
  \frac{(A+2)q^2}{2(A+1)}-{\cal E}}\right),
\label{tauncbar}
\end{eqnarray}
\begin{eqnarray}
K_{i=1,2}^\ell(q,p;{\cal E})&\equiv& G_i^\ell(q,p;{\cal
E})-\delta_{\ell 0}\; G_i^\ell(q,p;-\mu^2),
\label{ki} \\
\label{g1} G_i^\ell(q,p;{\cal
E})&=&\int_{-1}^{1}\hspace{-0.1cm}dy\frac{P_\ell(y)} {{\cal
E}-\frac{A+1}{A+A^{i-1}}q^2- \frac{A+1}{2A}p^2 - A^{1-i} {p q y}
+{\rm i}\epsilon}.
\end{eqnarray}
The Kronecker delta $\delta_{\ell 0}$ ($= 1$ for $\ell=0$ and =0 for
$\ell\ne 0$) is introduced in order to allow for finite results,
using a subtracted kernel, only for $\ell=0$ where such
regularization is necessary. The Thomas collapse is absent for
$\ell>0$, due to the centrifugal barrier, and such procedure is not
necessary. The subtraction energy $-\mu^2$ allows to fix the
two-neutron separation energy of $^{20}$C to its physical value
giving our renormalization procedure.

In this way, the Eq.~(\ref{Veq}) is  renormalized and the three-body
observables are completely defined by the two-body energy scales,
$\varepsilon_{nc}$ and $\varepsilon_{nn}$, and the energy of
$^{20}$C. The regularization scale $\mu^2$, used in the $s-$wave
[see Eq.~(\ref{ki})], is chosen to reproduce the three-body
ground-state energy of $^{20}$C  with a value of $-3.5$
MeV~\cite{audi}. The scaling function  for $s-$wave observables has
a limit cycle~\cite{braaten,mohr} evidenced when $\mu$ is let to be
infinity. We point out that a good description of this limit is
already reached in the first cycle~\cite{virtual}.

The numerical method to calculate the elastic scattering amplitude
relies on the use of an auxiliary function~\cite{adhprc81}, which is
a solution of an integral equation similar to the original one, but
with a kernel without the unitarity cut, due to a subtraction
procedure at the momentum $k$.  The scattering amplitude is
obtained by evaluating certain integrals over the auxiliary
function, when the two-body unitarity cut is introduced back. In the
present case, we have the following integral equation for the
auxiliary function $\Gamma$, and the corresponding solution for
$h_n^\ell(q;{\cal E})$:
\begin{eqnarray}
{\Gamma_n^\ell}(q,k;{\cal E})&=&{\cal V}^\ell(q,k;{\cal E})
+\frac{2}{\pi}\int_0^\infty\hspace{-0.1cm} dp \left[p^2{\cal
V}^\ell(q,p;{\cal E})- k^2{\cal V}^\ell(q,k;{\cal E})\right]
\frac{{\Gamma_n^\ell}(p,k;{\cal E})}{p^2-k^2},
\nonumber\\
{h_n^\ell}(q;{\cal E})&=&
\frac{{\Gamma_n^\ell}(q, k;{\cal E})}
{\displaystyle 1-\frac{2}{\pi}{k^2}\int_0^\infty dp
\;\frac{{\Gamma_n^\ell}(p,{ k};{\cal E})}{p^2-k^2-{\rm
i}\epsilon}} .\label{hna2}
\end{eqnarray}

For the on-shell scattering amplitude, we have
\begin{eqnarray}
&&h_n^\ell(k;{\cal E})= [ k\cot\delta_\ell - {\rm i}k]^{-1}
\label{a3} ,
 \end{eqnarray}
such that, from Eq.~(\ref{hna2}), we have
\begin{eqnarray}
&& k\cot\delta_\ell
= \frac{1}{{\Gamma_n^\ell}(k,k;{\cal E})} 
\left[1-
\frac{2}{\pi}k^2\int_0^\infty dp
\;\frac{{\Gamma_n^\ell}(p,k;{\cal E}) - 
{\Gamma_n^\ell}(k,k;{\cal E})}{p^2-k^2}\right] .  \nonumber
\end{eqnarray}
Note that the numerical stability and accuracy
of the results is delicate when an Efimov state is near the
scattering region. In this situation the method from Ref.~\cite{adhprc81} as outlined above, is far much accurate than the
use of contour deformation technique directly in Eq.~(\ref{Veq}).

To guide the discussion of the results of $ k\cot\delta^R_0$ for the
$n-^{19}$C elastic scattering from the numerical solution of 
Eqs.~(\ref{hna2}),  in analogy with the form given by Eq.~(\ref{kcotnd}),
we consider an equivalent low energy parametrization  of the effective
range expansion in terms of the kinetic energy $E_K$:
\begin{eqnarray}
k\cot\delta_0^{R}=\frac{-a^{-1}_{n-^{19}C} +\beta\; E_K  + \gamma\;
E_K^2}{1-E_K/E_0},
  \label{kcotd}
\end{eqnarray}
where $a_{n-^{19}C}$ is the ${n-^{19}C}$ scattering length, with
$\beta$ and $\gamma$ the effective  range parameters to be adjusted.
$E_0$ is the position of the pole with respect to the threshold for
elastic scattering.

\begin{figure}[tbh!]
\centerline{\epsfig{figure=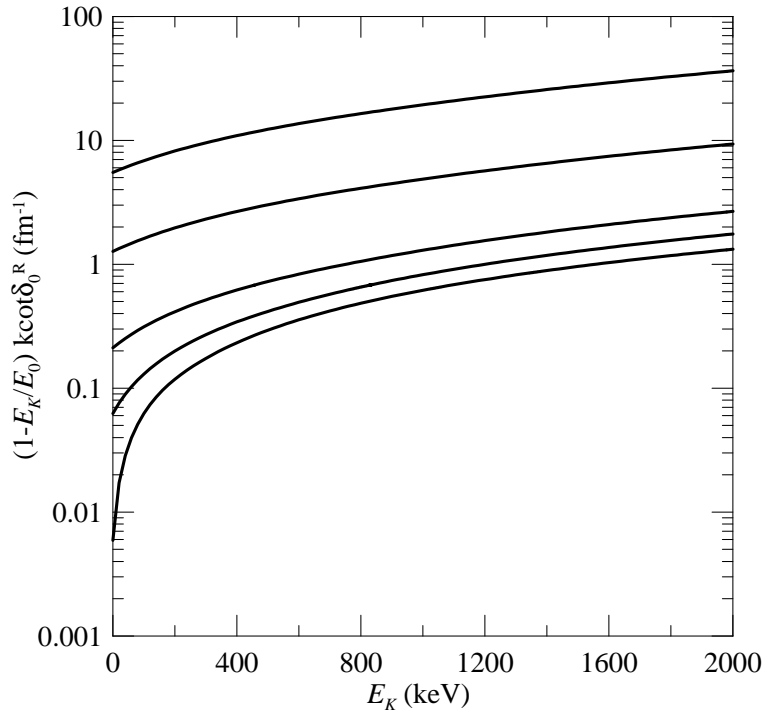,width=10cm}} 
\caption{
The function $(1-E_K/E_0) k\cot\delta_0^R$, where $E_0$ is the pole position, 
is given in terms of the CM kinetic energy $E_K$.
From bottom to top the curves correspond to the following $^{19}$C
binding energies (absolute values): 200, 400, 600, 800 and 850 keV.} \label{figkcot}
\end{figure}

\begin{figure}[tbh!]
\centerline{\epsfig{figure=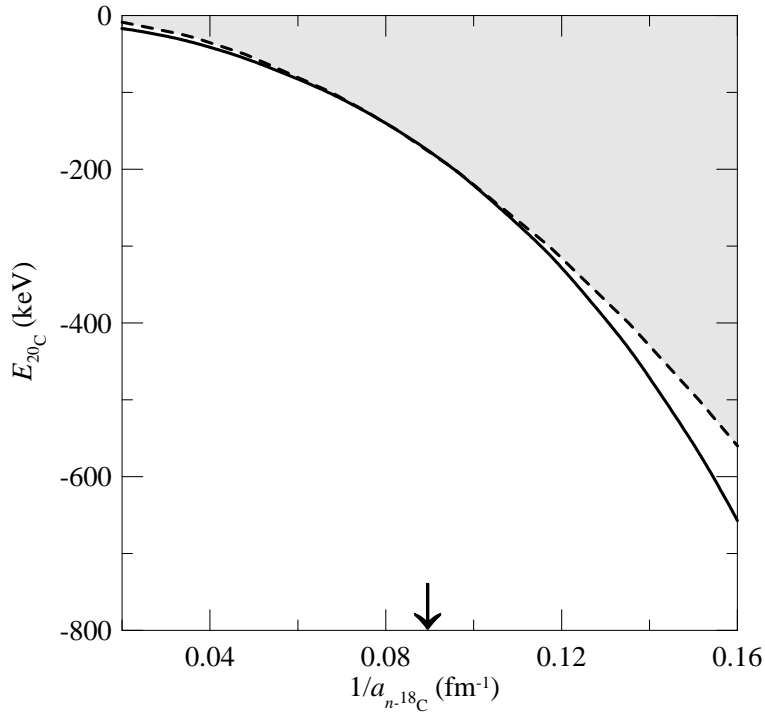,width=10cm}}
\caption{Energies of the first excited (bound or virtual) state of $^{20}$C are given as 
a function of the inverse of the $n-^{18}$C scattering length (solid line).
The dashed-line (where $E_{^{20}{\rm C}}$ = $E_{^{19}{\rm C}}$) splits the non-allowed region 
(shadowed area) from the allowed one. By increasing the energy of $^{19}$C (decreasing $a_{n-^{18}\rm{C}}$), 
the arrow indicates the value for which an excited Efimov state turns into a virtual state.  } \label{fige2e3}
\end{figure}

The numerical solution of Eq.~(\ref{hna2}) gives a zero in
$\delta_0^R$, as anticipated by our discussion. In order to present
smooth curves, in Fig.~\ref{figkcot} we show the results for 
$(1-E_K/E_0)k\cot\delta^{R}_0$ in terms of the CM kinetic energy $E_K$.
The calculations were performed for different values of $|E_{^{19}C}|$
below 850 keV. For $|E_{^{19}C}|>$ 170 keV, the $^{20}$C presents an Efimov 
virtual state~\cite{c20}, which can be seen in Fig.~\ref{fige2e3}, where it is 
plotted the $^{20}$C energies against the inverse of the $n-^{18}$C scattering lengths, 
$a_{n-^{18}{\rm C}}$. Below this value, the excited Efimov state turns out to
be bound.  The given {\it positive} $a_{n-^{18}{\rm C}}$ are related to 
the $^{19}$C binding energies by 
$E_{^{19}{\rm C}} = [\hbar^2/(2\mu_{n-^{18}{\rm C})}] (1/a_{n-^{18}{\rm C}}^2)$,
where $\mu_{n-^{18}{\rm C}}$ is the reduced mass for the system ${n-^{18}{\rm C}}$,
such that $(1/a_{n-^{18}{\rm C}}) \approx 0.00676 \sqrt{E_{^{19}{\rm C}}/{\rm keV}}$ fm$^{-1}$. 
The threshold for $E_{^{20}{\rm C}}$ is given by the dashed line.
The non-allowed area is the upper-right part of the figure (shadowed region). 
The arrow marks approximately the point where an excited $E_{^{20}{\rm C}}$ state 
(on the left) becomes a virtual state (on the right).
In table \ref{table1} we give the effective range parameters from a
fit of (\ref{kcotd}) to the results shown in Fig.~\ref{figkcot}.
\begin{table}[thb!]
  \caption{Effective range parameters in Eq.~(\ref{kcotd}) for different
$|E_{^{19}{\rm C}}|$ energies.}
\begin{tabular}{|ccccc|}
\hline
$|E_{^{19}{\rm C}}| $(keV)& $(a_{n-^{19}C})^{-1}$ (fm$^{-1}$) & $\beta$ (fm.keV)$^{-1}$ & $\gamma$ (fm.keV$^2$)$^{-1}$& $E_0$ (keV)   \\
\hline
200 & -0.591 10$^{-2}$  & 5.685 10$^{-4}$  & 4.673 10$^{-8}$& 1442.7\\
400 & -0.624 10$^{-1}$  & 6.743 10$^{-4}$  & 8.821 10$^{-8}$& 823.9\\
600 & -2.118 10$^{-1}$  & 9.337 10$^{-4}$  & 1.464 10$^{-7}$& 451.4\\
800 & -1.268            & 3.110 10$^{-3}$  & 4.424 10$^{-7}$& 115.0\\
850 & -5.510            & 1.201 10$^{-2}$  & 1.641 10$^{-6}$& 28.8\\
\hline
\end{tabular}
\label{table1}
 \end{table}
The $n-^{19}$C scattering length is indeed negative as the nearby
Efimov state is virtual, in agreement with the findings of Ref.~\cite{c20}. 
The Efimov excited state moves from a bound state (first energy sheet - 
left-hand side of the arrow of Fig.~\ref{fige2e3}) to a virtual state 
(second energy sheet - right-hand side of the arrow of Fig.~\ref{fige2e3}) 
through the elastic analytical cut, for $|E_{^{19}{\rm C}}|$ approximately 170 keV
($a_{n-^{18}{\rm C}}\approx 11.347$ fm)
In fact, for 200 keV, the scattering length is very large but still negative indicating 
the presence of a nearby pole of the S-matrix in the second energy sheet. We also observe the 
increase of the effective range parameters with the $^{19}$C energy.
With the $n-n$ energy fixed to $E_{nn}= -$143 keV and considering 
the experimental results for the neutron separation energy from $^{19}$C 
($-$160$\pm$110 in \cite{audi} and $-$530$\pm$130 keV in \cite{naka99}), the
existence of excited bound Efimov states in $n-n-^{18}$C is quite doubtful.
In order to have a bound excited state, in the present zero-range approach, 
we need $|E_{^{19}{\rm C}}| < $170 keV, which is excluded in \cite{naka99}. Only the lower 
limits (absolute values) given in \cite{audi} are allowing such possibility. However, we
should observe that range corrections will increase the allowed values of 
$|E_{^{19}{\rm C}}|$ for a bound $E_{^{20}{\rm C}}$ excited state~\cite{TFJ}.

\begin{figure}[tbh!]
{\epsfig{figure=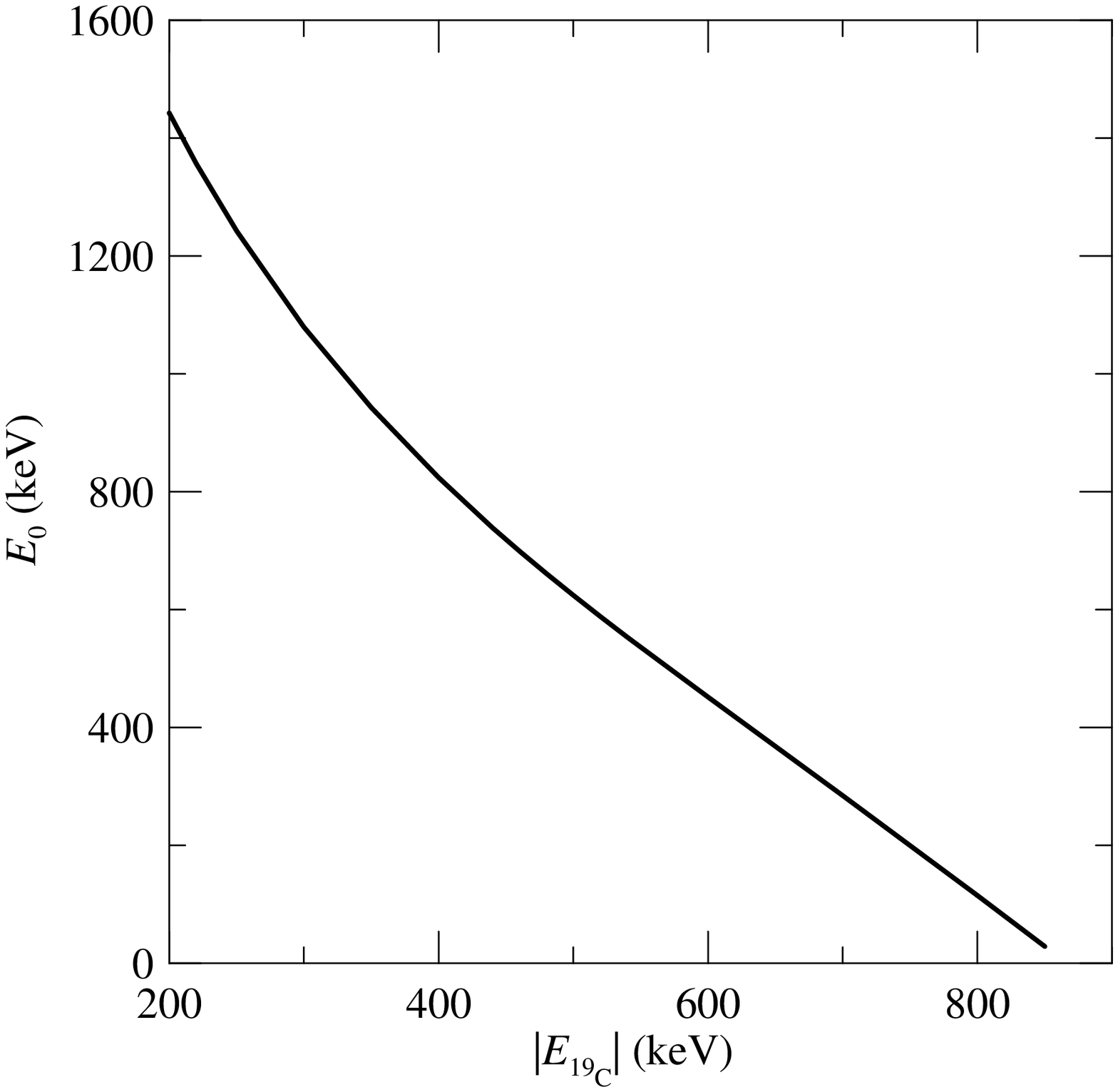,width=6.9cm}\epsfig{figure=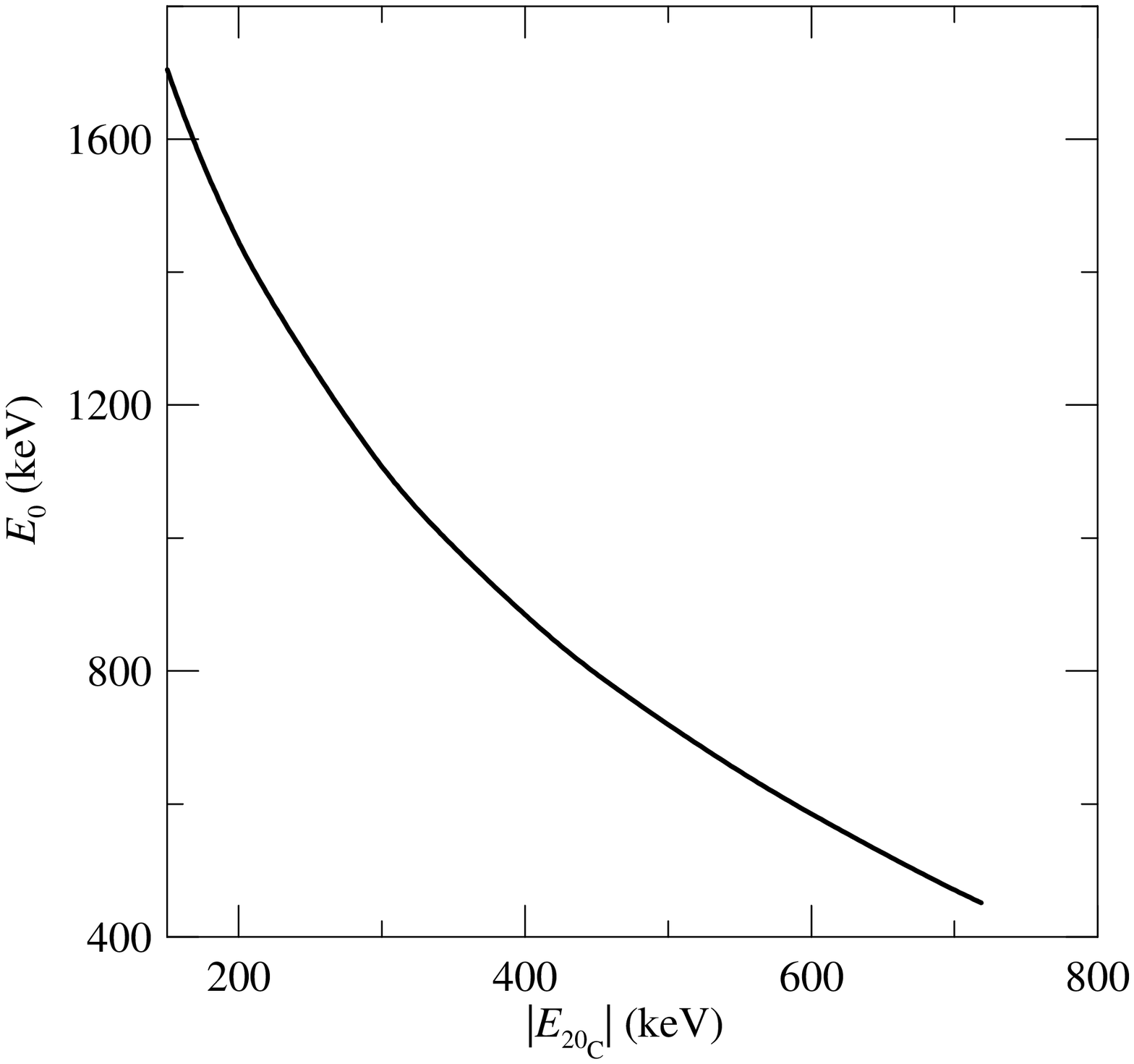,width=7.1cm}}
\caption{Position of the pole in $k\cot\delta_0^R$ is shown as a function of the 
absolute values of the $n-^{19}$C binding energy (l.h.s.); and 
excited bound/virtual $^{20}$C energy (r.h.s.).}\label{figpoloc19c20}
\end{figure}

The position of the pole in $ k\cot\delta_0^{R}$ is also shown in
Fig.~\ref{figpoloc19c20} as a function of $|E_{^{19}{\rm C}}|$
(l.h.s.) and the energy of the virtual $^{20}$C state (r.h.s.),
respectively. Moving towards  small values of $|E_{^{19}{\rm C}}|$
the pole moves to larger energies. While the increase of
$|E_{^{19}{\rm C}}|$ makes the scattering length goes towards zero
somewhat above 850 keV, as indicated by the l.h.s. of 
Fig.~\ref{figpoloc19c20}. The $s-$wave elastic cross-sections is damped
at low-energies for parameters around these values. For
$|E_{^{19}{\rm C}}|$ above 500 keV, the pole is below the breakup
threshold and the corresponding $s-$wave cross-section presents a
zero. For $|E_{^{19}{\rm C}}|$ below 500 keV, the pole is above the
breakup threshold where absorption occurs, and consequently the
$s-$wave cross-section has a minimum, not a zero, at the position of
the pole. Clearly, as the Efimov state moves deeper (see r.h.s. of
Fig.~\ref{figpoloc19c20}) in the second energy sheet, the pole
tends to zero energy. This behavior can be easily verified by using
the effective range expansion (\ref{kcotd}) and calculating the pole
of the scattering amplitude for the virtual state energy (see e.g.
\cite{adhikaritorr}).  As shown by our above numerical results, we
observe that in order to go from a zero in the scattering amplitude
to  the situation where one virtual Efimov state becomes bound, the
ratio of the two-body scattering lengths is given by
$\sqrt{170/850}\simeq 0.45$, which is quite close to our previous
estimative for $A=18$ ($a_0/a_B\simeq 0.42$),  as discussed after
Eq.~(\ref{bra3}).  Although the estimative is made using a
qualitative approach, it is indicative of the relation between the
physics of the Efimov effect and the zero in the scattering
amplitude. We should remark that a zero in the scattering length
does not imply, in general, the Efimov physics, while the opposite
is true, as we have substantiated by the qualitative and
quantitative calculations of the $n-^{19}$C elastic phase-shift.
\begin{figure}[tbh!]
\centerline{\epsfig{figure=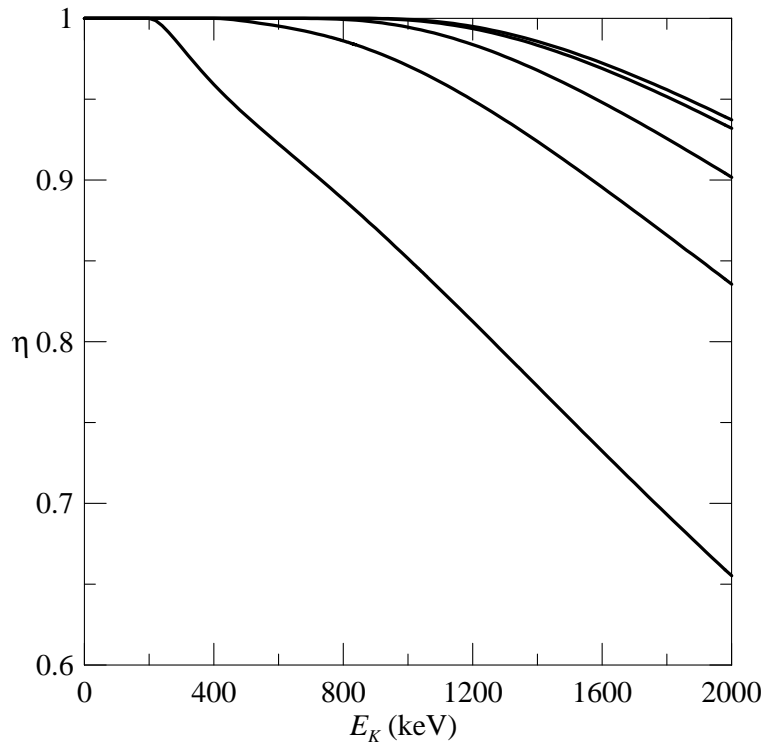,width=10cm}}
\caption{ $s-$wave
absorption parameter as a function of the CM kinetic energy $E_K$. 
From left to right the curves corresponds to the following $^{19}$C
energies: 200, 400, 600, 800 and 850 keV. }
\label{figabs}
\end{figure}
The absorption parameter $\eta=|e^{\imath \delta_0}|$ is shown in
Fig.~\ref{figabs}. It changes strongly with the binding energy of
$^{19}$C. We observe that the absorption increases naturally with
the size of $^{19}$C, as we see in the weakly bound case of 200 keV.

In conclusion, we studied the elastic scattering of a neutron on
$^{19}$C near the condition for an excited Efimov state of $^{20}$C.
In our work the energy of the ground state of $^{20}$C and the
neutron-neutron scattering length were fixed to the experimental
values. We allowed a change in the absolute value of the neutron 
binding energy of $^{19}$C up to 850 keV, in  order to include the
experimental range of values and seek for the interesting Efimov
related  physics in the $s-$wave elastic scattering amplitude. The
effective range expansion of the $s-$wave phase-shift presents a
low-energy pole, that moves toward zero energy as the binding of
the neutron in $^{19}$C increases and the virtual Efimov state goes
deeper in the second energy sheet. The parametrization of the
phase-shift was done by a simple analytical formula of the effective
range expansion with a pole, proposed long ago by Van Oers and
Seagrave\cite{vanoers} to fit the low-energy experimental data of
the doublet $s$-wave neutron-deuteron phase-shift. The three-body
$n-n-^{18}$C system shares with the trinucleon system the large
subsystem scattering lengths, the small binding energy of the ground
state and the proximity of a virtual  Efimov state to the scattering
region. These common features  implies that both systems show
analogous universal behaviors~\cite{bedaque-bira,fedorov02}, having qualitatively
properties independent on the mass ratios and described within zero
range models. Our study extends to the $n-^{19}$C elastic scattering
the general properties found in the neutron-deuteron system, in
particular the pole in the effective range expansion. Actually, it
should be of interest to extend the present analysis of the
low-energy elastic scattering to other possible two-body
configurations, with different mass relations within the three
particle system. Our results, obtained by using a renormalized
zero-range approach, which is valid in the limit of large two-particle 
scattering length, correspond essentially to the dominance of
the long range $\rho^{-2}$ interaction, responsible for the Efimov
effect, Thomas collapse, as well as the zero in the $n-^{19}$C
scattering length. It is expected that a more realistic approach for 
the neutron-core interaction~\cite{TFJ}, due to range effects,
will increase the possible region of $^{19}$C binding energies 
for the existence of an excited bound state (moving the arrow mark 
in the Fig.~\ref{fige2e3} to the right).

Loosely bound neutron-rich nuclei near the drip-line, such as the carbon 
and oxygen isotopes with two-neutron halos, are actually the more 
promising nuclear systems that are being intensively investigated~\cite{petra} 
in order to reach a better understanding on the nuclear forces properties and interactions.
From another side, the experiments with loosely bound two-neutron halo nuclei 
can also give relevant informations on the few-body scales and universality. 
In this respect, as discussed in Ref.~\cite{HS} for carbon isotopes, 
the three-body $n-n-^{20}$C system, is probably a more favorable system 
to study low-energy three-body properties, considering that the two-neutron
binding energy in $^{22}$C is close to zero, implying in a very large 
neutron halo. This is a Borromean case with large possibilities for
Efimov resonant states.
Another relevant source of information on universality and the 
corresponding dominance of few scales at very 
low-energy actually can also be obtained from ultracold atomic 
physics experiments.
The possibility of different two-body scattering lengths  can be
realized in trapped ultracold atomic systems near a Feshbach
resonance. In this situation the environment of ultracold atomic
traps allows to follow the scaling of three-body observables with
two-body scattering lengths. The atom-dimer scattering length
($a_{AD}$) can change from large and positive to negative and then
to zero moving the two-atom scattering length (c.f.
Eq.~(\ref{bra1})), allowing to control the effective atom-dimer
interaction near the Efimov limit. Consequently, the stability of
the atom-dimer condensate should be sensitive to an Efimov state
that passes from bound to virtual, i.e., the effective interaction
proportional  to $a_{AD}$ changes  from strongly repulsive to
attractive, and then to zero by lowering the value of $a$. In
particular, when $a_{AD}=0$, the atom and dimer condensates are
invisible to each other, i.e., they decouple. It is conceivable that
in an experiment would be possible to dial the coupling between the
atom and dimer condensates, and thus controlling the phases of the
condensed gas mixture. Interesting phenomena in the condensate due
to Efimov states near the scattering threshold have already been
discussed in Refs.~\cite{newphases}. The mixed atom-dimer phases
near the Efimov limit add to the rich physics of trapped atoms more
possibilities to be tested experimentally. In fact, the scaling laws
of few-body observables near the Efimov limit is under active
experimental investigation, even in more complex situations like
that of the four-boson system \cite{grimm08}, where the quest for
new scales is under debate \cite{ejl06,hammer07}. The recent
measurements of the dimer-dimer recombination in ultracold traps
near the Feshbach resonance \cite{grimm08} and Efimov limit, asks for
further theoretical analysis in order to extend the concepts used
here at the three-body level to the four-boson scattering.

We thank Funda\c c\~ao de Amparo \`a Pesquisa do Estado de S\~ao
Paulo and Conselho Nacional de Desenvolvimento Cient\'\i fico e
Tecnol\'ogico for partial support.

\end{document}